\documentclass[journal]{IEEEtran}
\IEEEoverridecommandlockouts
\usepackage{cite}
\usepackage{amsmath,amssymb,amsfonts}
\usepackage{algorithmic}
\usepackage{graphicx}
\usepackage{textcomp}
\usepackage{booktabs}
\usepackage{nicefrac}
\usepackage[pdftex,dvipsnames]{xcolor}
\def\BibTeX{{\rm B\kern-.05em{\sc i\kern-.025em b}\kern-.08em
    T\kern-.1667em\lower.7ex\hbox{E}\kern-.125emX}}

\usepackage{lipsum}
\usepackage{xargs} 
\usepackage{url}

\newcommand\todoM[1]{{}}

\begin{document}

\title{On Optimal Battery Sizing for Households Participating in Demand-Side Management Schemes %
\thanks{The authors want to thank Hamad Bin Khalifa University and the Doctoral Training Alliance (DTA) Energy for their support.\newline M.~Pilz is with the School of Computer Science \& Mathematics at Kingston University London in Kingston upon Thames, UK.\newline O. Ellabban is with Iberdrola Innovation Middle East, Qatar Science \& Technology Park, Doha, Qatar.\newline L. Al-Fagih is with the Division of Engineering Management and Decision Sciences in the College of Science and Engineering at Hamad Bin Khalifa University, Qatar.}%
}

\author{Matthias Pilz, Omar Ellabban, and Luluwah Al-Fagih}

\maketitle

\begin{abstract}
	The smart grid with its two-way communication and bi-directional power layers is a cornerstone in the combat against global warming. It allows for the large scale adoption of distributed (individually-owned) renewable energy resources such as solar photovoltaic systems. Their intermittency poses a threat to the stability of the grid which can be addressed by the introduction of energy storage systems. Determining the optimal capacity of a battery has been an active area of research in recent years. In this research an in-depth analysis of the relation between optimal capacity, and demand and generation patterns is performed for households taking part in a community-wide demand-side management scheme. The scheme is based on a non-cooperative dynamic game approach in which participants compete for the lowest electricity bill by scheduling their energy storage systems. The results are evaluated based on self-consumption, the peak-to-average ratio of the aggregated load, and potential cost reductions. Furthermore, the difference between individually-owned batteries to a centralised community energy storage system serving the whole community is investigated.
\end{abstract}


\section{Introduction}
	Global average temperatures are rising dramatically (2016 being the warmest year ever recorded~\cite{NASA2019a}), causing a noticeable increase in natural disasters and environmental issues~\cite{NASA2019}. Thus it is imperative to investigate approaches that reduce greenhouse gas emissions and slow down climate change. Instead of burning fossil fuels to satisfy our energy needs, humans should make use of renewable energy resources such as solar and wind, which have a smaller carbon footprint. In order to guarantee a stable power grid, demand and generation have to be balanced at all times. This makes the integration of renewable resources a challenging task due to their intermittent nature. The advent of the smart grid, a technologically advanced power grid, is a possible solution to this problem. It combines the legacy power grid with a communication layer, effectively connecting all the grid participants. Through this additional infrastructure energy consumption can be managed. 
	
	More specifically, in this research, the functionality to exchange data between individual households is being used to schedule energy storage installations such that the grid stability is guaranteed even though a considerable amount of demand is served from solar power generation facilities. A key element to achieve a high self-consumption rate of solar energy, i.e.~the ratio between the consumed solar energy to the actual demand, is the utilisation of energy storage. Various research studies are concerned with energy storage management~\cite{Soliman2014,Nguyen2015,Li2018,Luthander2015,Li2014,Pilz2018}.

	Luthander \textit{et al.}~\cite{Luthander2015} present a case study of 21 Swedish households with a focus on comparing individually-owned batteries to a centralised storage solution. In order to reach a certain level of self-consumption the centralised storage capacity is considerably smaller than the aggregated capacity of individually-owned batteries.
	The study in~\cite{Li2014} is concerned with optimising the usage of a given photovoltaic-battery system. It investigates a number of different optimisation objectives and shows how these affect the eventual charging patterns of a household for two exemplary days. 
	In contrast to their approach, Reference~\cite{Pilz2018} makes use of a game-theoretic approach in which households schedule their individually-owned batteries with the goal to minimise their respective electricity bills. They perform simulations over the period of an entire year to allow for statistical analysis of the results.  
	
	One interesting question is: \textit{What is the optimal capacity of a battery?}~\cite{Huang2018,Khalilpour2016,Talent2018}. Reference~\cite{Talent2018} focuses on the influence of different tariff schemes on the optimal battery size, whereas~\cite{Khalilpour2016} develops a decision-making tool which supports users that are investing in photovoltaic and battery systems. Recently, Huang et al.~\cite{Huang2018} developed an algorithm to determine the optimal size of a battery with respect to the achievable self-consumption. This research builds on their approach and develops a deeper understanding of the relation between demand and generation patterns, and the optimal battery capacity.

	The main contributions of this research are as follows:
		\begin{itemize}
			\item[(1)] Based on seasonal and yearly simulations of households with real consumption and generation data, this research provides an in-depth insight on how optimal sizing of batteries depends not only on aggregated statistics but also on the specific temporal patterns that characterise individual households.
			\item[(2)] Two different battery scheduling algorithms are compared in terms of three metrics: (i) Self-consumption of solar energy, (ii) Peak-to-average ratio of the aggregated load as an indicator for grid stability, and (iii) Potential cost reductions due to the introduction of electricity storage systems.
			\item[(3)] This research compares the optimal sizing for a centralised storage facility with individually-owned batteries and analyses their effect on the same metrics as mentioned above.
		\end{itemize}

	This paper is organised as follows. In Section~\ref{sec:system}, the neighbourhood model is presented. This includes a brief summary of the models for the storage and generation systems, general definitions of demand and load, as well as the role of the utility company. Furthermore, the underlying battery scheduling game is explained. Section~\ref{sec:results} introduces the dataset and simulation parameters before summarising, analysing, and discussing the results. The paper is concluded in Section~\ref{sec:conclusion}.

\section{System Model and Scheduling Objectives}
\label{sec:system}
	In this section, the model for the community and its participants is briefly summarised. For a more detailed description of the setup the reader is kindly referred to~\cite{Pilz2018}.
	\subsection{Neighbourhood}
	\label{subsec:neighbourhood}
		Consider a neighbourhood of $M$ households that is modeled as a set $\mathcal{M}$. Each of the households $m\in\mathcal{M}$ is equipped with a smart meter, an individually-owned battery, and a photovoltaic (PV) system that converts solar energy into electricity. The smart meters have the capability to measure consumption and generation data over equidistant time intervals $t\in\mathcal{T}$. A day is split into $T$ intervals. Furthermore, the smart meters are able to exchange data through wireless communications.

	\subsection{Households}
	\label{subsec:households}
		A detailed battery model is employed. It includes the charging and discharging efficiency ($\eta_+,\ \eta_-$) of the storage device, self-discharging at a rate of $\bar{\rho}$ in case the battery is idle, and also limits on how much can be charged or discharged ($\rho_+,\ \rho_-$) during a particular interval $t\in\mathcal{T}$. Furthermore, the conversion efficiency $\eta_{\text{inv}}$ of the DC/AC power electronics converter is considered. 
	The main equations characterising the battery model are as follows, more details are presented in~\cite{Pilz2018}:
		\begin{subequations}
			\begin{align}
				SOC^{t+1} = SOC^t + \eta_{\text{inv}}\eta_+ a_+^t &\ \  \text{ charging} \\
				SOC^{t+1} = SOC^t + \nicefrac{a_-^t}{\eta_{\text{inv}}\eta_-}  &\ \  \text{ discharging} \\
				0 \leq a_+^t < \rho_+ &\ \  \text{ charging rate}\\
				\rho_- < a_-^t \leq 0 &\ \  \text{ discharging rate}\\
				\text{SOC}_{\min} \leq \text{SOC}^t \leq \text{SOC}_{\max} &\ \  \text{ limited SOC}\ ,	
			\end{align}
		\end{subequations}
		
		where SOC is the state-of-charge of the battery, $a_+^t$ and $a_-^t$ are the charging and discharging amount in interval $t$, respectively, and $\text{SOC}_{\min}$ and $\text{SOC}_{\max}$ give the lower and upper bound of the SOC.	
		
		Each household is equipped with a solar PV system for local consumption. The PV systems vary in size according to the dataset described in Section~\ref{subsec:data}. The solar energy generated locally $w^t_m$ is always taken into account before making a scheduling decision for a specific interval $t$. This means that if available, the energy from the PV system is used to fulfil the demand $\bar{d}^t_m$ of a household. Furthermore, if the solar PV generation exceeds the demand it is stored in the local battery (if possible). Note that charging the battery from the solar PV system does not require DC/AC conversion, whereas it needs to be converted to AC before it is being used to fulfil demand.
		
		Therefore, the net-demand of a household $m\in\mathcal{M}$ at time interval $t\in\mathcal{T}$ can be written as
		\begin{equation}
			d^t_m = \bar{d}^t_m - w^t_m\ .
		\end{equation}
		Then the load $l^t$ on the electricity grid, i.e.~the amount of electricity that is provided by the utility company, can be written as
		\begin{equation}
			l^t_m = d^t_m + a^t_m\ .
		\end{equation}

	\subsection{Utility Company}
	\label{subsec:utilityCompany}
		All the community households are supplied by one network distribution company, which incentivises each household individually to reduce their energy consumption during peak hours. A billing strategy which calculates a unit energy price per interval $t$ based on the aggregated load of all customers is given by
		\begin{equation}
			p^t = c_2 \cdot y^t + c_1 \cdot y^t + c_0\ ,
			\label{eqn:price}
		\end{equation}
		where $y^t$ is the aggregated load of all users at interval $t$ and $c_2 > 0$, $c_1,c_0 \geq 0$ are constants. As a consequence the energy bill for a particular day of an individual household can be calculated as:
		\begin{equation}
			b_m = \sum_t l^t_m \cdot p^t\ .
			\label{eqn:bill}
		\end{equation}

	\subsection{Game Scheduling and Self-Consumption Constraint}
	\label{subsec:GTSvsGTSWC}
		The batteries of the households are scheduled based on a dynamic game which is played between the individual households. The objective of the players/households is to minimise their individual electricity bill \eqref{eqn:bill}. They act rationally and in a selfish manner. In the following section, two approaches are differentiated. The first approach is identical to the one proposed by~\cite{Pilz2018} and will be called ``Game-Theoretic Scheduling'' (GTS). The game is played for an upcoming day based on forecasts for demand and generation. 
		
		The second approach indroduces an additional constraint to the GTS. Whenever the renewable PV generated energy is expected to be higher than the demand for an upcoming interval, charging the battery from the grid is prohibited. The idea behind this is to maximise the self-consumption rate of the PV system. This approach will be referred to as ``Game-Theoretic Scheduling with Constraint'' (GTSWC).

\section{Simulations with the Ausgrid Dataset}
\label{sec:results}
	In this section, the Ausgrid dataset and further simulation details are presented. Then the evaluation metrics are defined. Eventually, the simulation results are analysed and discussed. 
	
	\subsection{Data, Simulation Details, and Metrics}
	\label{subsec:data}
		For all the following simulations the real-world ``Ausgrid'' dataset~\cite{Ausgrid2019} is being used. This dataset has been collected half hourly, i.e.~$T=48$ intervals, from 300 individual homes in the Ausgrid's electricity network (New South Wales, Australia) over three years (2010-2013). All of the homes are equipped with solar PV systems between 1.05 kWp and 9.99 kWp. The period between July 2010 and June 2011 has been split into four seasons as shown in Tab.~\ref{tab:seasons}. 
		\begin{table}
			\centering
			\caption{\textit{Season Definitions}.}
			\label{tab:seasons}
			\begin{tabular}{ll}
			\toprule
			season & period \\\midrule
			winter & $01/07/2010 - 28/09/2010$ \\
			spring & $01/10/2010 - 29/12/2010$ \\ 
			summer & $01/01/2011 - 31/03/2011$ \\
			autumn & $01/04/2011 - 29/06/2011$ \\\bottomrule			
			\end{tabular}
		\end{table}
		This is done such that each individual season spans exactly 90 days. A clean version of the Ausgrid dataset, which contains $M=54$ households is considered in this research. Please refer to~\cite[Chapter 2]{Ratnam2016} and~\cite{Ellabban2019} for a thorough analysis of the complete dataset. The consumption and generation data are used as an input to the day-ahead scheduling mechanism described in Section~\ref{subsec:GTSvsGTSWC}. This means they are treated as forecasted data and throughout this study no forecasting errors are considered.
		
		The optimal battery size for each household is determined following a process reported by Huang \textit{et al.}~\cite{Huang2018}. As there are two scheduling approaches (GTS and GTSWC), this is done twice for each season and also independently for an entire year resulting in lists of optimal battery capacities for each household. After that one run for each approach and season in which the households are equipped with their individually determined optimal battery size has been performed. For comparison of the outcomes, the following three metrics are being used: (i) The percentage increase in self-consumption by the introduction of the battery, (ii) The cost reduction due to the introduction of the battery according to the billing function shown in Section~\ref{subsec:utilityCompany}, and (iii) The peak-to-average ratio (PAR) of the aggregated load of all the households, i.e.
		\begin{equation}
			\text{PAR} = T \cdot \frac{\max_m \sum_t l_m^t}{\sum_n \sum_{\tau} l_n^{\tau}}\ .
			\label{eqn:PAR}
		\end{equation}

	\subsection{Optimal Battery Sizing Results and Discussion}
\begin{figure}
			\centering
			\includegraphics[width=\linewidth]{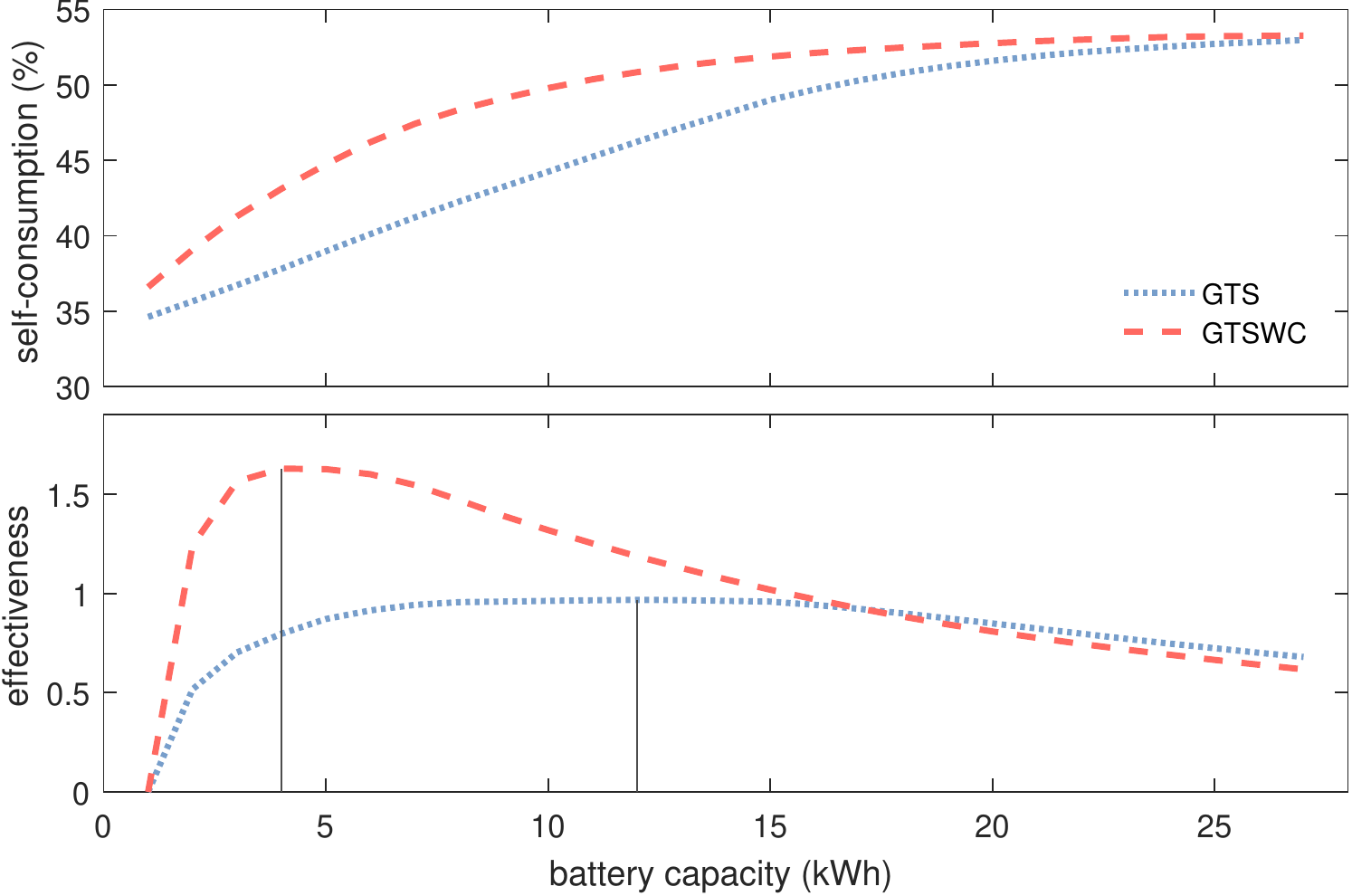}
			\caption{\textit{Optimal sizing considerations}. The self-consumption and the resulting effectiveness of an exemplary household are plotted over the battery size. The two vertical lines indicate the maximum of the effectiveness for the GTS and GTSWC approach and therefore the optimal size of the energy storage installation, respectively. }
			\label{fig:sizingProcedure}
		\end{figure}

		In order to determine the optimal battery size, the process described in~\cite{Huang2018} has been followed. To do so, simulations are performed with different battery sizes for each household (per season and yearly) using both scheduling approaches. Battery capacities are in the range between 1.0 kWh and 27.0 kWh. The upper limit would equal an installation of two Tesla Powerwall2 batteries~\cite{Tesla2017}. For each set of parameters, the `effectiveness' of the electricity storage is computed. The effectiveness is defined by the notion of how much the self-consumption of a household is increased per kWh of installed capacity:
		\begin{equation}
			\text{effectiveness} = \frac{sc_n - sc_1}{n}\ ,
			\label{eqn:effectiveness}
		\end{equation} 
		where $sc_n$ is the self-consumption achieved by utilising storage of size $n$ kWh.
		The maximum of this effectiveness is the sought after optimal battery size. An example for these steps is shown in Fig.~\ref{fig:sizingProcedure} for a randomly selected house and season.
				
\begin{figure*}
			\centering
			\includegraphics[width=\textwidth]{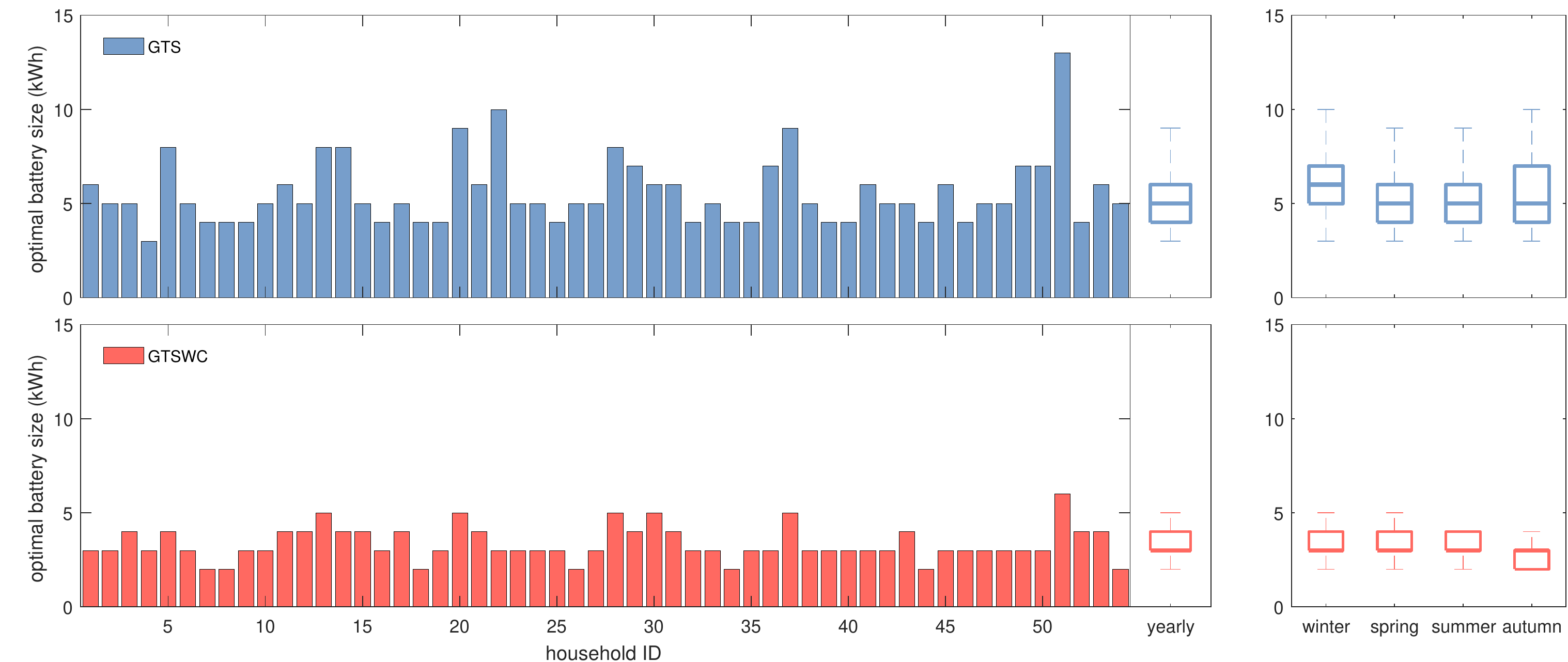}
			\caption{\textit{Optimal battery sizes}. The results are obtained through a process as described in~\cite{Huang2018}.Battery capacities between 1 and 27 kWh were analysed. The optimal battery sizes for the individual households from simulation runs over the period of an entire year are reported. Furthermore, statistical results for these simulations as well as independent seasonal simulations are shown.}
			\label{fig:optSizes}
		\end{figure*}

		The optimal battery size for each player over the course of an entire year for the two approaches (game-theoretic scheduling with and without self-consumption constraint, cf.~Section~\ref{subsec:GTSvsGTSWC}) is shown in Fig.~\ref{fig:optSizes}. Furthermore, Fig.~\ref{fig:optSizes} also reports the average results per season over all the 54 investigated households.

		Overall, the optimal size for the GTSWC scenario does not exceed the GTS optimal battery size for any player and season. Houshold 4 shows the smallest difference between the two scenarios. All the capacities are the same except for summer where they differ by 1 kWh. The largest difference between the optimal battery size as determined for the two approaches of a particular season is found in household 14 (summer). Here the difference is 8 kWh (cf.~Fig.~\ref{fig:sizingProcedure}). The largest difference between the optimal battery size for two seasons and the same approach is seen in household 52 (winter: 11 kWh, summer: 3 kWh). In Section~\ref{sec:batteryAnalysis}, these households are investigated in particular to understand how their battery usage patterns lead to the respective results.\\
				
		\subsubsection{Self-Consumption}
		\label{subsub:selfConsumption}
\begin{figure}
				\centering
				\includegraphics[width=\linewidth]{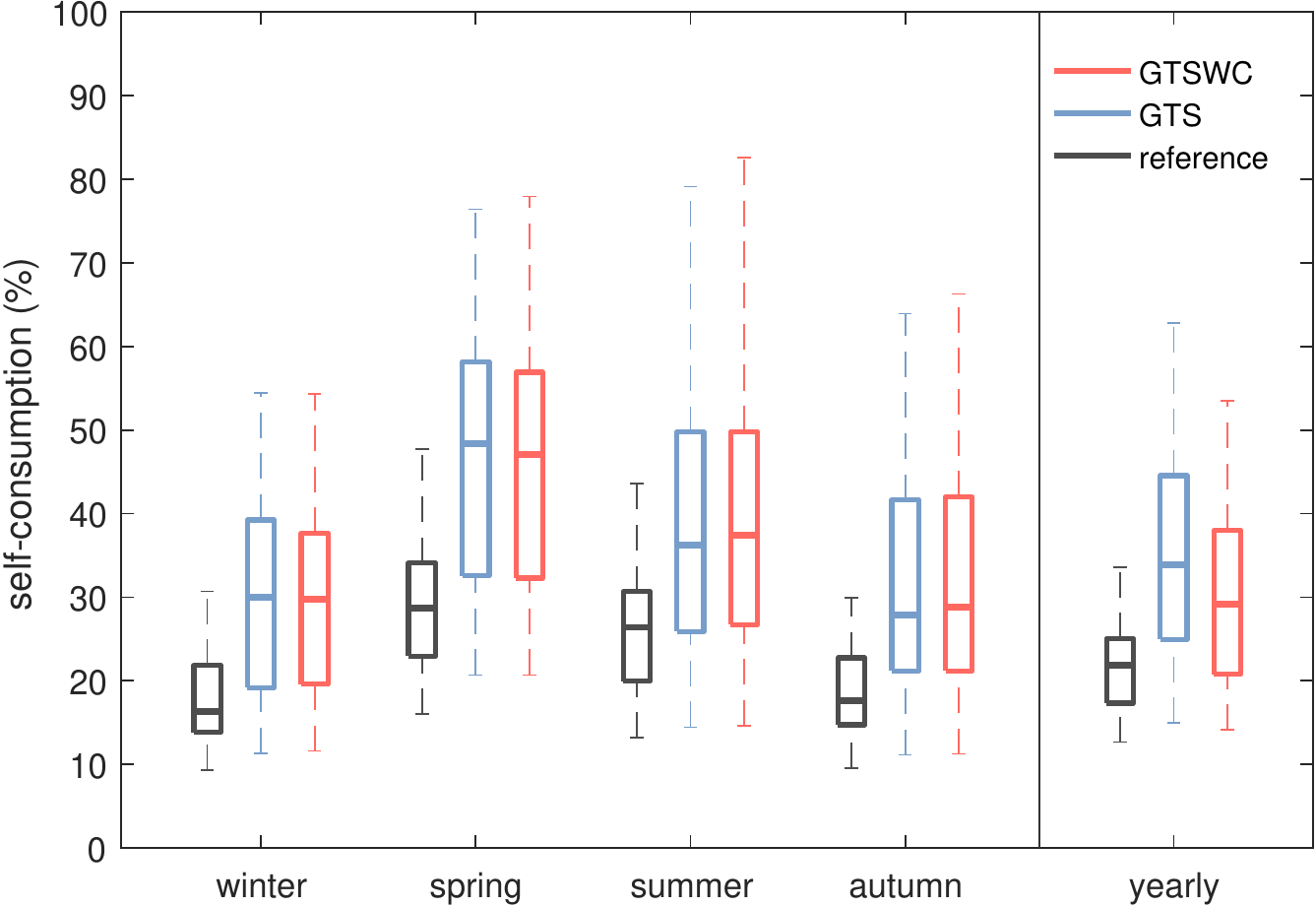}
				\caption{\textit{Self-consumption analysis}. Statistical results for the self-consumption rates are shown for all seasons and an entire year. For each period, the reference case in which no storage is available is compared with a configuration that includes the optimally sized batteries for each individual household for both the GTS and the GTSWC approach.} 
				\label{fig:selfConsumption}
			\end{figure}

			The solar PV self-consumption rate of a household is defined as the ratio between the solar energy being used and its demand. This includes a direct part which is consumed immediately and an indirect part used to charge the battery when the PV system generation exceeds the local demand. In the following, the increase in self-consumption due to the introduction of an optimally sized battery for both the GTS and GTSWC scenario is analysed. The seasonal results for the self-consumption can be found in Fig.~\ref{fig:selfConsumption}. Explicit improvements are reported in Tab.~\ref{tab:selfConsumption}.
			\begin{table}
				\centering
				\caption{\textit{Self-consumption improvements}. The median improvement (over all households) of the self-consumption due to the introduction of optimally sized batteries is shown. The simulations for each column were performed independently.}
				\label{tab:selfConsumption}
				\begin{tabular}{p{1.1cm}|p{0.9cm}p{0.9cm}p{0.9cm}p{0.9cm}|p{0.9cm}}
					\toprule
				 	& winter & spring & summer & autumn & yearly \\\midrule
				 	GTS & $11.4\%$ & $17.8\%$ & $10.0\%$ & $11.3\%$ & $11.6\%$  \\
				 	GTSWC & $12.1\%$ & $17.0\%$ & $10.7\%$ & $12.2\%$ & $12.5\%$  \\\bottomrule
				\end{tabular}
		\end{table}
		It becomes clear that even with different optimal battery sizes for the GTS and GTSWC approach the median improvement in self-consumption is similar. The result was to be expected as the additional constraint in GTSWC is particularly designed to place further emphasis on the increase of self-consumption. The spread around the median self-consumption approximately doubles when comparing the results with the reference case in which no batteries are present. This is due to the fact that some households benefit more than others from the introduction of a battery. There are many factors that play a role for this such as the aggregated solar production, the aggregated demand, and also the temporal patterns of production and demand. For example: Household 14 improves its self-consumption by $12.2\%$ during the summer, while household 26 (with similar aggregated consumption and PV peak production) improves its self-consumption by $1.5\%$. Household 13, which has less aggregated demand than the two houses mentioned before and higher aggregated solar production, improves its self-consumption by $37.9\%$. A more detailed analysis of these households and how these differences are related to their demand patterns will be analysed in Section~\ref{sec:batteryAnalysis}. In general\footnote{Explicit results for this statement are not shown due to lack of space.}, households that gain considerably at GTS also do so at GTSWC and vice versa. The average absolute difference of the self-consumption improvements between GTS and GTSWC for each household individually is $<1.4\%$.\\
		\subsubsection{PAR values}
		\label{subsub:PAR}
\begin{figure}
				\centering
				\includegraphics[width=\linewidth]{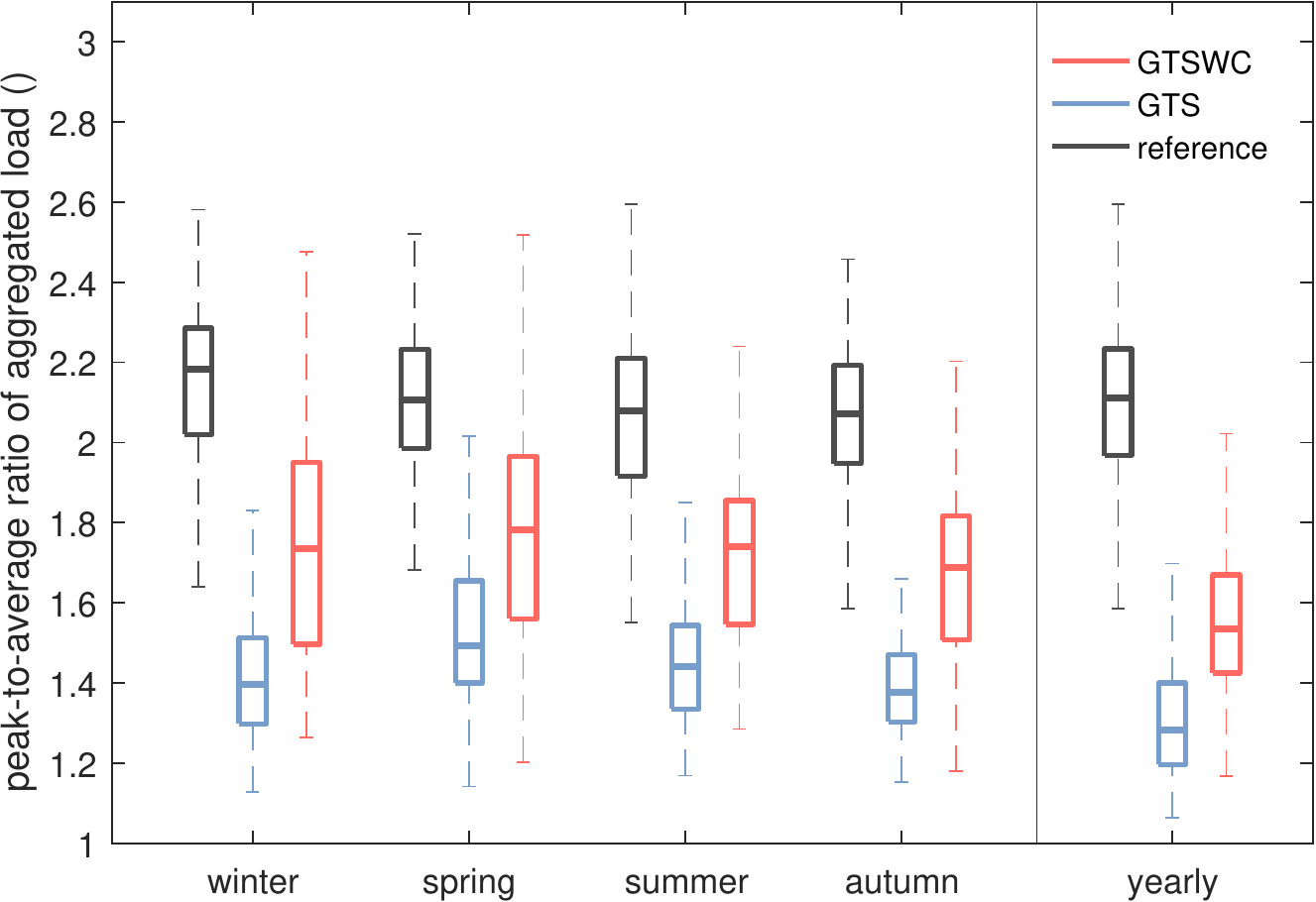}
				\caption{\textit{Peak-to-average ratio (PAR) of the aggregated load.} A statistical analysis of the achieved daily PAR values over the respective seasons is shown. For each period, the reference case in which no storage is available is compared with a configuration that includes the optimally sized batteries for each individual household for both the GTS and the GTSWC approach.}
				\label{fig:PAR}
			\end{figure}

			The peak-to-average ratio (PAR) of the aggregated load~\eqref{eqn:PAR} is an indicator for the stability of the grid~\cite{Bayram2014}. A value close to 1.0, i.e.~a flat load profile, is preferred by a utility company as this allows them to save investment costs for fast-ramping energy production installations. PAR values are calculated for a period of one day. A statistical analysis for the 90 days that comprise each season is shown in Fig.~\ref{fig:PAR}.
			Overall, considerable improvements of the PAR value are achieved. The GTS approach leads to better PAR reductions than the GTSWC approach in both the median values and also the smaller spread around these. \\

		\subsubsection{Cost Reduction}
\begin{figure}
				\centering
				\includegraphics[width=\linewidth]{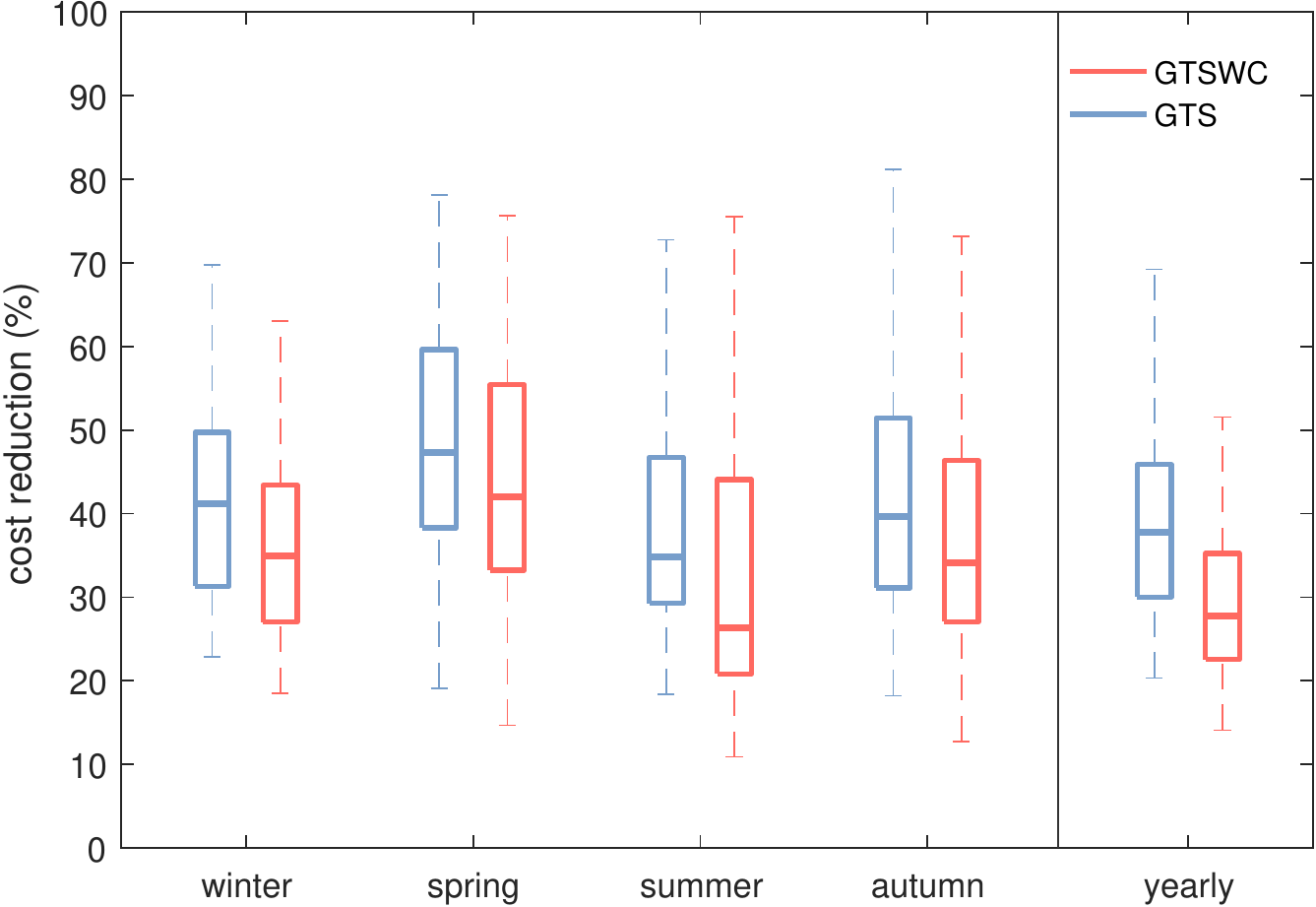}
				\caption{\textit{Cost reductions}. Statistical analysis of the amount of savings from the electricity bill over various billing periods is presented for the GTS and the GTSWC approach. The calculation of the unit energy price depends on the aggregated load as introduced in~\eqref{eqn:price}.}
				\label{fig:costReduction}
			\end{figure}

			As seen in Section~\ref{subsec:utilityCompany}, the cost function~\eqref{eqn:price} depends on the aggregated load. Thus the price per unit of electricity changes for each half hour interval. When calculating the overall bill for each household with and without battery, it can be observed that it is decreasing in both approaches. The relative cost reduction of the electricity bill due to the introduction of an optimally sized battery is shown in Fig.~\ref{fig:costReduction}.
			Overall, the introduction of energy storage leads to a considerable amount of savings from the electricity bill in both cases (GTS and GTSWC). The increase of  self-consumption can directly be translated in a decrease of energy requested from the grid which in turn decreases the bill. As seen in the previous section (cf.~Section~\ref{subsub:selfConsumption}), the achieved improvements in self-consumption are similar for the two approaches. This means this fact alone cannot explain the higher savings from GTS compared to GTSWC. The second factor that plays a role is the more effective PAR reduction observed for the GTS approach (cf.~Section~\ref{subsub:PAR}). Due to the quadratic relation between the aggregated load and the price per unit of electricity, consumption during peak times is billed highly. The spread around the median values for both approaches is similar.
	\newpage		
	\subsection{Analysis of Battery Usage Patterns}
	\label{sec:batteryAnalysis}
\begin{figure*}
			\centering
			\includegraphics[width=\textwidth]{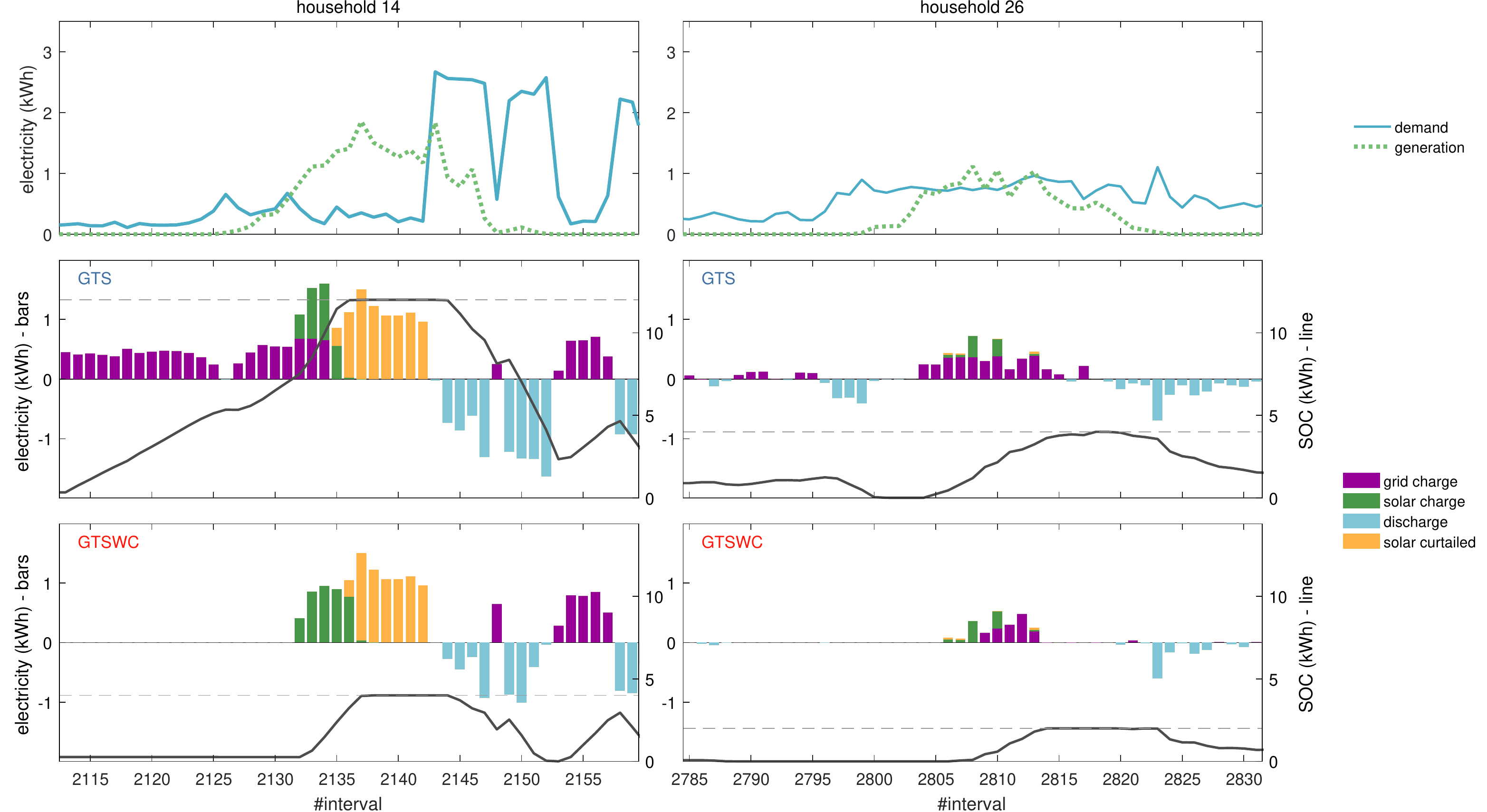}		
			\caption{\textit{Demand, generation, battery usage}. The demand and generation profiles of household 14 (left) and household 26 (right) for representative days are shown. Furthermore, the specific battery usage based on the GTS and GTSWC approach are presented. For these four plots, the left hand axes represent the electricity values for the bars, while the right hand axes indicate the state of charge (SOC) of the respective battery. The dotted line indicates the optimal battery size for the repsective household. \newline For this particular day household 14 improves their self-consumption by $6.0\%$ / $10.4\%$ through the GTS / GTSWC approach, respectively. Household 26 improves their self-consumption by $2.7\%$ for both approaches.}
			\label{fig:analysis_twoPlayers}
		\end{figure*}

	While the aggregated demand of a household and the size of their installed solar panel can give a rough estimate for the optimal battery size, it remains important to look at the actual demand and generation patterns. In Section~\ref{subsub:selfConsumption} it was visible that two households with similar aggregated demand and peak PV output benefited differently from their storage installation. In order to understand this difference, Fig.~\ref{fig:analysis_twoPlayers} shows the demand and generation profile for a randomly chosen day together with the detailed battery usage for these two households.
		The demand of household 14 is low during the time when solar is available and peaks shortly afterwards, whereas the demand of household 26 is rather evenly distributed throughout the day. Household 26 is a prime example of a user that has a high percentage of non-curtailed solar energy even without a battery installation and cannot gain much through the utilisation of storage. Consequently, the optimal battery sizing algorithm determines a below-average optimal storage capacity for this household. In contrast to this, the battery of household 14 is optimally sized at an above-average capacity. Without storage a lot of the solar energy is curtailed due to the lack of demand at the particular time of generation. Since there is a peak in demand in the later hours, the self-consumption can be increased due to the storage capability.
		
		The left-hand plots in Fig.~\ref{fig:analysis_twoPlayers} also give insight into the differences between the GTS and GTSWC approach. During the fist half of the day, the GTS algorithm charges the battery from the grid, whereas the GTSWC anticipates the solar generation and thus restricts charging from the grid. The first two peaks in demand (cf.~between interval 2144 and 2152) can then be met by previously saved electricity. In anticipation of another peak in demand at the end of the day, both algorithms charge the battery and are able to flatten the load curve considerably. It becomes clear that because no more solar production is to be expected during this time there are no constraints on charging the battery from the grid and both algorithms behave similarly.

		\begin{figure*}
			\centering
			\includegraphics[width=\linewidth]{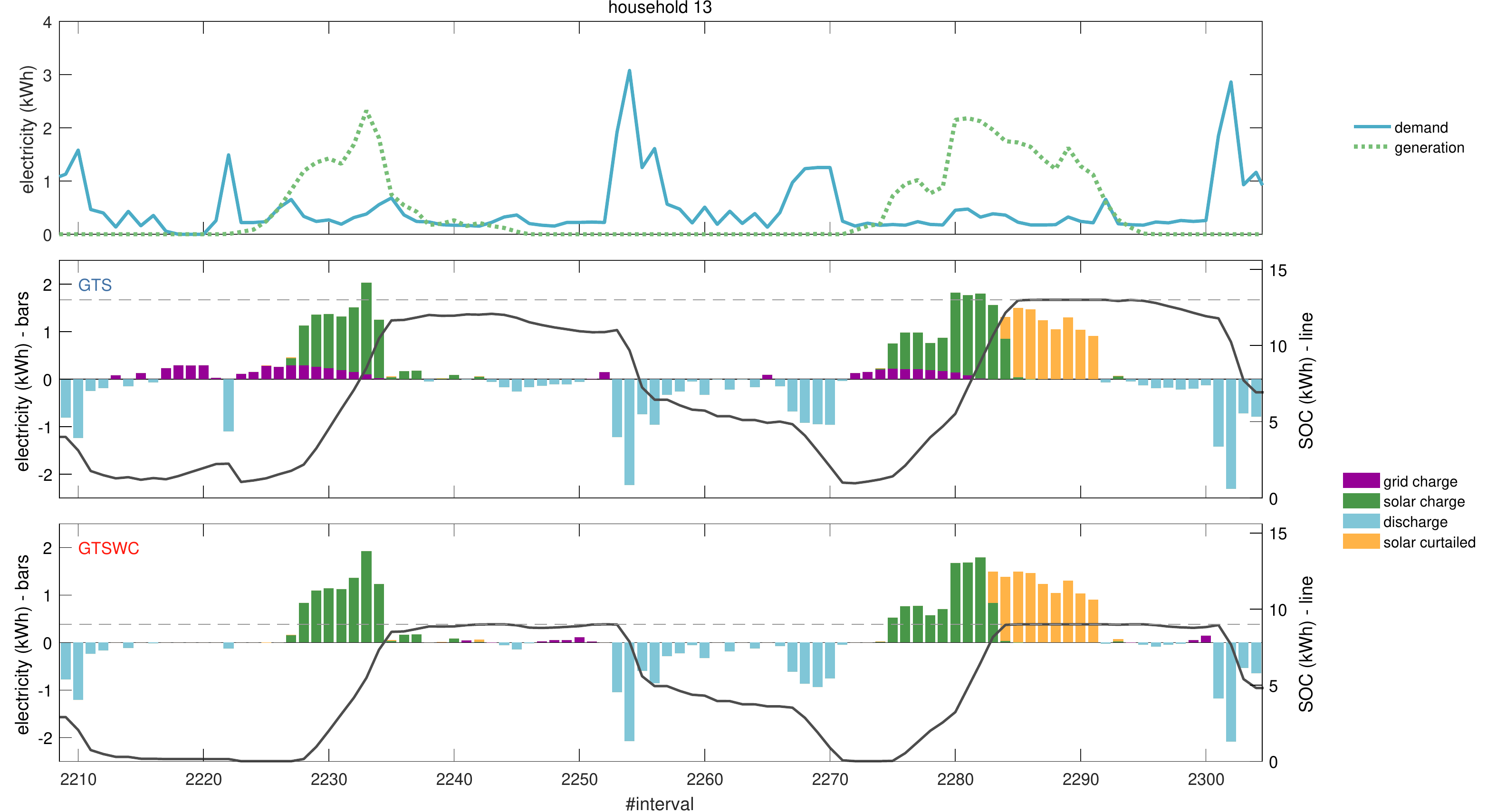}
			\caption{\textit{Demand, generation, battery usage}. The demand and generation profiles of household 13 for two representative days are shown. Furthermore, the specific battery usage based on the GTS and GTSWC approach are presented. For the lower two plots, the left hand axes represent the electricity values for the bars, while the right hand axes indicate the state of charge (SOC) of the respective battery.}
			\label{fig:analysis_13_winterOld}
		\end{figure*}
		Fig.~\ref{fig:analysis_13_winterOld} shows the demand and generation pattern together with the battery usage for two consecutive days of household 13. This household was chosen as it has the highest benefit during this particular season from installing an optimal battery size.
		A similar profile for the demand and generation as seen for household 14 in Fig.~\ref{fig:analysis_twoPlayers} can be observed. The even higher improvement in self-consumption for this case stem from the more pronounced asynchronisation between solar generation and actual demand. Also, this household is equipped with a bigger solar panel.		
		
\begin{figure}
			\centering
			\includegraphics[width=\linewidth]{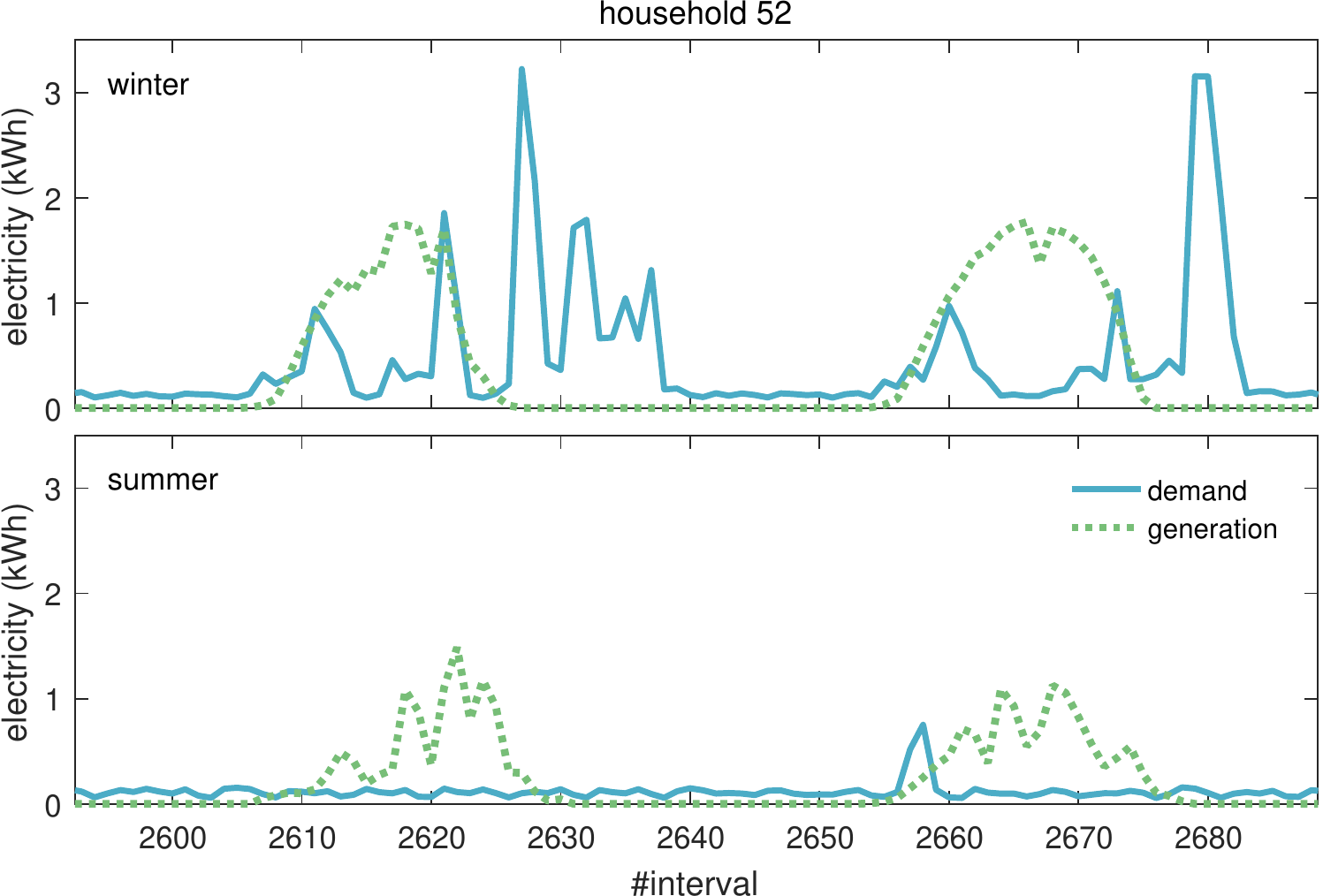}
			\caption{\textit{Demand and generation for two seasons}. The demand and PV generation of household 52 for two consecutive representative days of winter 2010 and summer 2010 are shown.}
			\label{fig:player52}
		\end{figure}

		This section is concluded by analysing the demand and generation profile (Fig.~\ref{fig:player52}) for the household that showed the biggest difference in optimal battery size between two seasons, i.e.~household 52. For winter 2010 the optimal capacity is determined to be 11 kWh while in summer 2011 it is 3 kWh. 		
		\newpage
	\subsection{Centralised vs. Decentralised Storage Systems}
	\label{subsec:community}
		In all the previous simulations, each household was in possession of an individual battery of different size. Within this section, a scenario that has a single battery to serve the community is investigated. For a reasonable comparison, the efficiency of the battery and the DC/AC power electronics converter equal the values used before. 
		\begin{figure*}
			\centering
			\includegraphics[width=\textwidth]{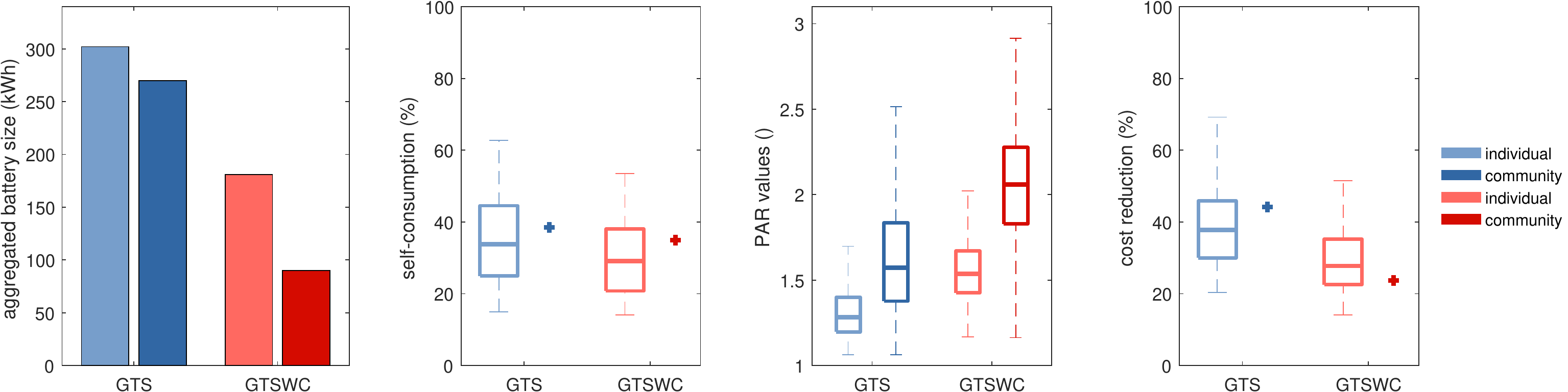}
			\caption{\textit{Comparison between centralised and decentralised approach}. The aggregated optimal battery sizes for the GTS and GTSWC approach in case of a single centralised battery and individually-owned decentralised batteries are shown. Furthermore the three metrics: self-consumption, peak-to-average ratio (PAR) of the aggregated load, and cost reduction are investigated for simulations based on these optimally sized storage installations. All simulations are performed over the period of an entire year, i.e.~winter 2010 to autumn 2011.}
			\label{fig:community}
		\end{figure*}
		Furthermore, the maximal charging and discharging rates were scaled up by the number of households. Firstly, full-year simulations with battery sizes varying between 10 kWh and 370 kWh were performed. Following the optimal sizing procedure by~\cite{Huang2018}, the optimal battery capacity for both the GTS and GTSWC approach were calculated to be 270 kWh and 90 kWh, respectively. For these optimal sizes, the self-consumption, the peak-to-average ratio (PAR) of the aggregated load of all households, and the cost reduction according to the pricing function~\eqref{eqn:bill} are analysed. The results are shown together with the respective results from yearly simulations of individually-owned batteries in Fig.~\ref{fig:community}.
		
		The centralised optimal battery sizes are approximately $10\%$ and $50\%$ smaller than the aggregated capacities of the decentralised batteries for the GTS and GTSWC approach, respectively. This is in agreement with a previous study reported by Luthander \textit{et al}.~\cite{Luthander2015}. In Section~\ref{sec:batteryAnalysis}, it was shown that in case of asynchronous demand and generation profiles, a large battery is most beneficial, while in the opposite case a small battery is sufficient. When looking at the centralised battery, note that it is scheduled according to the aggregated demand and generation of all the households. An averaging effect for the demand profiles occurs, which makes the asynchronous case less likely and eventually leads to a smaller optimal storage capacity. 
		The PV self-consumption reaches a comparable level to the decentralised simulations. Compared to the median self-consumption of all the households, a scenario with a centralised battery improves the self-consumption by approximately $5\%$ for both the GTS and GTSWC approach.
		When analysing the daily PAR values, it becomes clear that the community batteries perform worse both in terms of the achieved median values and also the spread around it. From the utility companies' perspective this is a unfavourable result. Their most desirable objective is to reduce the PAR value as it guarantees grid stability and financial benefits in the long run. 
		
		The right-most panel in Fig.~\ref{fig:community} shows the results for the cost reduction for both approaches comparing the centralised and decentralised neighbourhoods. For the GTS approach the centralised community achieves an approximately $5\%$ higher cost reduction, while for the GTSWC approach the cost reduction is reduced by approximately $5\%$ compared to the median cost reduction of all the households with individually-owned batteries. Both results for the centralised battery are within the interquartile range of the respective analysis for the decentralised system.

\section{Conclusions}
\label{sec:conclusion}
	In this paper, a community of households that take part in a demand-side management scheme is analysed. The focus was to gain deeper understanding of optimal battery sizing. Both the characteristics that lead to the optimal battery size determination as well as the effect this optimal size has on solar photovoltaic (PV) self-consumption ratio, grid stability/security, and cost reductions for the users has been investigated. A key insight is that the temporal patterns of consumption and generation impact the battery sizing critically. This means battery sizing which is soley based on aggregated data might lead to unfavourable results. Households which benefit most from installing a energy storage system are those where the peak-production and peak-consumption is asynchronous, i.e.~during different intervals of the day.
	
	Furthermore, two different approaches for the demand-side management scheme were compared. Game-theoretic scheduling (GTS) is based on the ideas presented in~\cite{Pilz2018}. Here the main objective of the individual households is to minimise their electricity bills. The second approach introduced an additional constraint to the GTS which puts PV self-consumption before the minimisation of the costs. As a result it lead to considerably smaller optimal battery sizes. The drawback of the more constrained approach are the larger peak-to-average ratio of the aggregated load, i.e.~higher costs for the utility company to guarantee stability of the grid. In terms of costs a trade-off is achieved: On the one hand, the initial investments are smaller for GTSWC due to the smaller battery sizes. On the other hand, the cost reduction off the electricity bill are less beneficial.
	
	The final part of the paper compared individually-owned batteries with a scenario that includes a utility sized centralised storage system. The optimal battery size determined for the centralised system is smaller due to less pronounced asynchrony of the aggregated demand to the solar PV production. 
		
\section{Acknowledgements}
The authors would like to thank Jean-Christophe Nebel for useful comments and valuable discussions. 

\bibliographystyle{IEEEtran}

\end{document}